\documentclass[english,preprint]{aastex}
\usepackage[T1]{fontenc}
\usepackage[latin1]{inputenc}
\usepackage{babel}
\usepackage{graphics}

\makeatletter

\providecommand{\LyX}{L\kern-.1667em\lower.25em\hbox{Y}\kern-.125emX\@}

\makeatother
\begin{document}

\title{Discovery of X-Ray Pulsations from the Compact Central Source in
the Supernova Remnant 3C~58}

\author{Stephen S. Murray, Patrick O. Slane, Fredrick D. Seward, Scott M.
Ransom}

\affil{Harvard-Smithsonian Center for Astrophysics, Cambridge, MA 02138}

\email{ssm@head-cfa.harvard.edu}

\author{Bryan M. Gaensler%
\footnote{Hubble Fellow
}}

\affil{Center for Space Research, Massachusetts Institute of Technology,
Cambridge, MA 02139}

\date{08 August 2001}

\keywords{ISM: individual (3C~58), pulsars: individual (PSR~J0205+6449),
stars:neutron, supernova remnants, X-rays:general}

\begin{abstract}
We report on high time and spatial resolution observations of the
supernova remnant 3C~58 using the High Resolution Camera (HRC) on
the \emph{Chandra X-ray Observatory}. These data show a point-like
central source, from which we detect 65.68-ms pulsations at \( 6.7\sigma  \)
significance. We interpret these pulsations as corresponding to a
young rotation-powered pulsar, PSR~J0205+6449, which is associated
with and which powers 3C~58. Analysis of archival \emph{RXTE} data
from three years earlier confirms these pulsations, and allows us
to determine a spin-down rate of \( \dot{P}=1.93\times 10^{-13} \)~s/s.
Assuming a magnetic dipole model for PSR~J0205+6449, we infer a surface
magnetic field of \( 3.6\times 10^{12} \)~G. The characteristic
age for this pulsar is 5400~yr, indicating either that 3C~58 was
not the supernova of 1181~CE, or that the pulsar's initial spin period
was \( \sim 60 \)~ms.
\end{abstract}

\section{Introduction \label{sect:intro}}

An isolated pulsar within a supernova remnant presents an opportunity
to compare the derived characteristics of both, particularly the ages.
3C~58 is of particular interest since it is generally accepted as
being the remnant of SN1181~CE (\cite{cla77}) and, therefore, has
a known age of 820 years. The classic best-loved example of a pulsar/SNR
association is the Crab Nebula, with characteristic pulsar age of
1250 years, close to the historical age of 946 years. Previous analysis
of observations of 3C~58 using Einstein and ROSAT (\cite{bec82,hel95})
suggested that there is a compact object at the center of the SNR.
The authors note that while 3C~58 is a young Crab-like remnant, its
radio luminosity is a factor of \( \sim 10 \) less than the Crab,
while its X-ray luminosity is \( \sim 2000 \) times below that of
Crab. This implies that the spin-down luminosity of any pulsar powering
3C~58 is substantially lower than that of the Crab, and that there
must be significant differences in the mechanisms that channel the
spin-down energy into X-ray emission. 

Using the ROSAT HRI data, and extracting photon events from this compact
object, \cite{hel95} failed to detect any periodic signal from this
putative pulsar. The Cycle 1 HRC GTO program included a search for
pulsations in a series of SNRs containing synchrotron nebulae.,with
3C~58 as the first target. It was selected as the most likely to
contain an X-ray pulsar based on the earlier Einstein (\cite{bec82})
and ROSAT(\cite{hel95}) observations. ASCA measurements showed that
the power-law spectrum steepens with distance from the center of the
nebula (\cite{tor00}), increasing the expectation that a pulsar is
powering the nebula.

In this paper we report on the results from two \emph{Chandra X-ray
Observatory} (CXO) HRC observations that were obtained for 3C~58
during Cycle 1 (1999-2000). One used the HRC-I as part of the original
observing program, and the second used the HRC-S in a special mode
designed to overcome a timing problem with HRC that was discovered
after the Chandra launch (\cite{mur01}). The goal of the observations
was to search for pulsations from the central source, and to measure
the spatial structure of the central region. We also report on a re-analysis
of \emph{Rossi X-ray Timing Explorer} (RXTE) observations that were
made in 1997 using a new suite of analysis tools developed for radio
pulsar searches and now applied to X-ray data (\cite{ran01}).

\section{Observations with the HRC}

\subsection{HRC-I Observation\label{sec:obsid00129}}

\begin{figure}[h]
{\centering \resizebox*{1\textwidth}{!}{\includegraphics{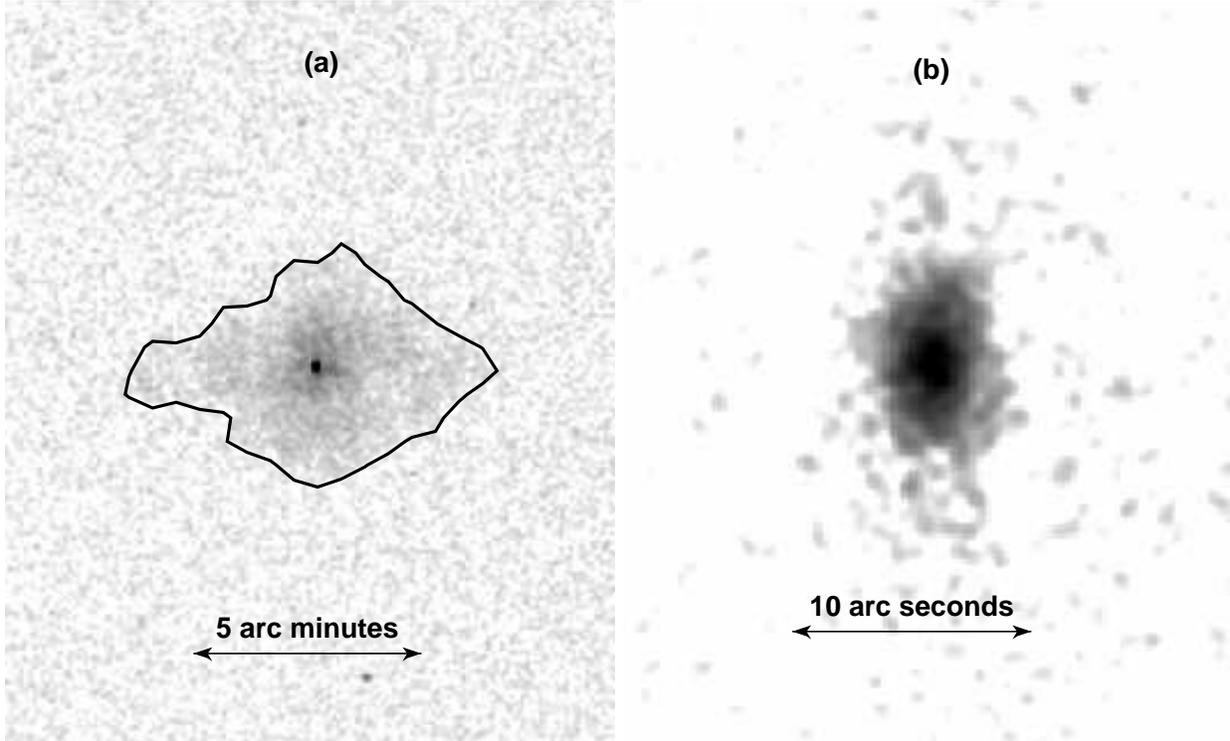}} \par}

\caption{{\footnotesize The HRC-I image of 3C~58 (OBSID 00129). The left
panel (a) shows the image binned at 16 HRC pixels (2.11 arc seconds)
per display pixel, and smoothed using a Gaussian with a 2 arc second
sigma. The contour shown is taken as the approximate limit of the
X-ray nebula and is used to estimate the total X-ray luminosity. The
image on the right (b) is the central region, binned at 1 HRC pixel
(0.1318 arc seconds) and smoothed using a Gaussian with a 0.25 arc
second sigma.} \label{fig:hrcimage}}
\end{figure}

The first HRC observation of 3C~58 was OBSID~00129. It was carried
out on 30, November 1999 and was done with HRC-I for 29 ksec. The
image obtained is shown in Figure~\ref{fig:hrcimage}. In panel (a),
the HRC data is binned to about 2.1 arc second/pixel (16 HRC pixels
per display pixel), and smoothed using a Gaussian with a 2 arc second
sigma. The contour shown corresponds to the X-ray isophote where the
nebula merges with the detector background (\( \sim 1.2 \) counts
per display pixel). The contour is about 8 arc minutes long and 5
arc minutes wide. Panel (b) shows the central part of the image. Here
the data are binned 0.1318 arc second/pixel, and then smoothed using
a Gaussian with a 0.25 arc second sigma. It is evident from these
Chandra images that the compact source and its extent first reported
by \cite{hel95} is real.

Integrating the X-ray emission over the entire nebula (as defined
by the contour) gives an estimate of its luminosity. There are 37883
total counts inside this contour, and after background subtraction
a net of 18620 counts in the 29 ksec observation time. Using the nebula
average best fit photon power-law spectral index \( \alpha =1.9 \),
and the column density \( N_{H}=3\times 10^{21}cm^{-2} \), (\cite{hel95,tor00})
we find \( L_{neb}=2.9\times 10^{34}erg/s \) (0.08 -10 keV), taking
the distance to 3C~58 to be 2.6 kpc (\cite{gre82})). The uncertainty
in luminosity due to various systematic effects such as the spatial
limit of the nebula and uncertainties in the spectral parameters is
estimated at about 20\%.

\subsection{HRC-S Observation\label{sect:obsid01848}}

The second HRC observation of 3C~58 used HRC-S in imaging mode to
allow accurate event timing measurements. In this 33 ksec observation
(OBSID~01848), taken 23 December, 2000 (MJD=51901.33), the same large
scale spatial properties were found for the source as described above.
However, in this observation the point source was located on-axis
resulting in better image quality on the sub-arc second scale. (In
the HRC-I observation the point source was about 2 arc minutes off-axis.)

\subsubsection{Two Dimensional Spatial Model of the Central Region\label{sec:2d}}

\begin{figure}[h]
{\centering \resizebox*{0.9\textwidth}{!}{\includegraphics{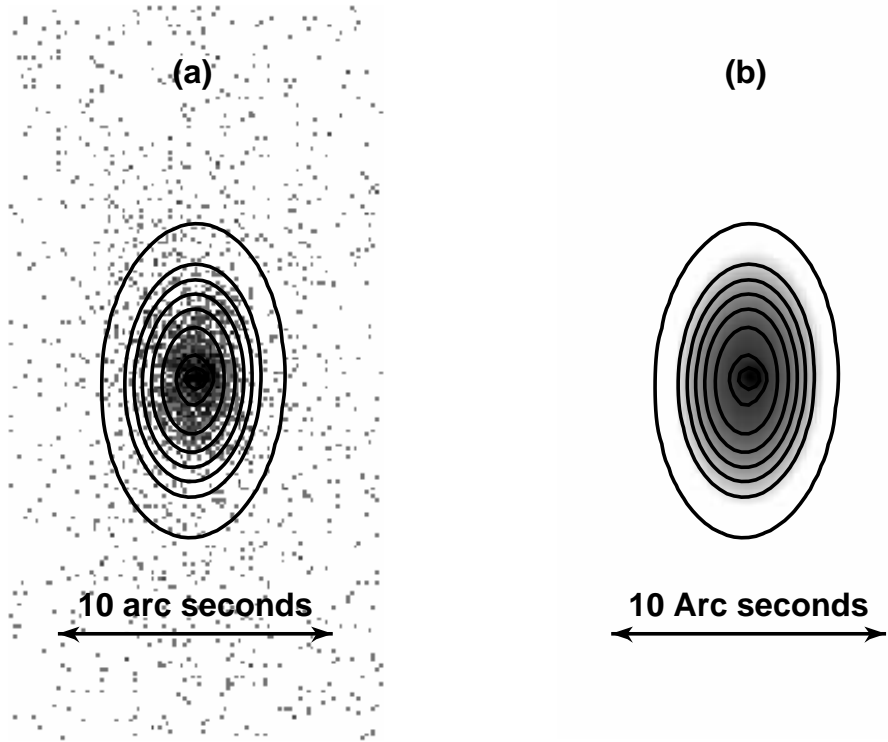}} \par}

\caption{{\footnotesize Central region of 3C~58 showing the compact source
and its extended nebula at 0.131 arc seconds per pixel (a) and a 2D
model consisting of a Gaussian approximation to a point-like source
plus an elliptical Gaussian representing the extended component (b).
The contours plotted are factor of two steps in surface brightness
starting at 0.039 cts\protect\( \: \protect \)pixel\protect\( ^{-1}\protect \).
For reference, the model contours are also plotted over the raw image
(a).} \label{fig:center}}
\end{figure}

The smoothed image of the central region of 3C~58 shown in the inset
of Figure~\ref{fig:hrcimage}b suggests that there is a point-like
source embedded in an extended region that is elongated along the
North-South direction. We have fit the central region of 3C~58 to
a simple 2D spatial model (using the CIAO Sherpa tools) consisting
of the sum of two 2D Gaussian distributions (one circularly symmetric
to simulate a point source, and one with ellipticity to simulate the
extended component), and a constant background. The results of the
spatial fit are shown in Figure~\ref{fig:center}, and summarized
in Table~\ref{tab:spatial}. From Table~\ref{tab:spatial} we note
that the first Gaussian component is consistent with a point source,
and that the extended emission is characterized by a FWHM of about
a 3 arc second extent along its major axis and about 1.3 arc seconds
along the minor axis.

{
\begin{table}[h]
{\centering \begin{tabular}{|c|c|c|}
\hline 
\multicolumn{3}{|c|}{Table~\ref{tab:spatial}~2D Spatial Model Results}\\
\hline
\hline 
Parameter&
Value&
Uncertainty\\
\hline
\hline 
g1.fwhm&
4.28&
\( \pm 0.48 \)\\
\hline 
g1.ampl&
7.32&
\( \pm 1.21 \)\\
\hline 
g2.fwhm&
25.78&
\( \pm 0.94 \)\\
\hline 
g2.ampl&
3.01&
\( \pm 0.18 \)\\
\hline
g2.ellip&
0.42&
\( \pm 0.03 \)\\
\hline
c3.c0&
0.038&
\( \pm 0.013 \)\\
\hline
\end{tabular}\par}

{\raggedright {\footnotesize \vspace*{0.1in}}\footnotesize \par}

{\raggedright {\footnotesize Table~\ref{tab:spatial}: The model
consists of the sum of the following components. A 2D Gaussian (g1)
with circular symmetry representing the point like source contribution,
a 2D Gaussian (g2) with elliptical symmetry representing the elongated
extended emission around the point-like source, and a constant 2D
background component (c3). We fixed the location of the Gaussian functions
at the centroid of the emission, and allowed the position angle and
ellipticity of the extended Gaussian to be free parameters. The FWHM
(in units of 0.131 arc seconds per pixel), the amplitude ( in counts/pixel
at the peak) for the Gaussian components, and the background amplitude
(counts/pixel) were also free parameters of the fit.}
\rule{6.5in}{1pt}\label{tab:spatial}\par}
\end{table}
\par}

The X-ray emission associated with the point source and its extended
sub-nebula is obtained by integrating the flux within the outer contour
shown in Figure~\ref{fig:center}. This is an isophote that is just
above the mean background (0.039 counts/pixel), and we have 2064 net
counts. Converting to luminosity, we use a photon power-law index
\( \alpha =2.0 \) for just the central region of 3C~58 (\cite{tor00}),
a column density of \( N_{H}=3\times 10^{21}cm^{-2} \), and a distance
to 3C~58 pc 2.6 kpc (\cite{gre82}) , to obtain \( L_{x-central}=2.84\times 10^{33}erg\: s^{-1}(0.08-10\: keV). \)
Using the model parameters, we estimate the point source contribution
to be 150 net counts. Converting to luminosity we obtain \( L_{x-point}=2.06\times 10^{32} \)
erg\( \:  \)s\( ^{-1} \)

\subsubsection{Timing Analysis}

In a \( \sim 1 \) arc second radius region centered on the point
source (RA=\( 02^{h}05^{m}37^{s}.8 \), DEC=\( +64^{\circ }49^{'}41^{''} \)
J2000), 744 photons were extracted from the image. The event times
(provided in terrestrial time by the CXC) were corrected to time at
the solar system barycenter using the definitive Chandra geocentric
spacecraft ephemeris. The standard CIAO axBary tool (using the JPL
DE450 solar system ephemeris) provided by the CXC was used to produce
a serial (time ordered) list of events.

\begin{figure}[h]
{\centering \includegraphics{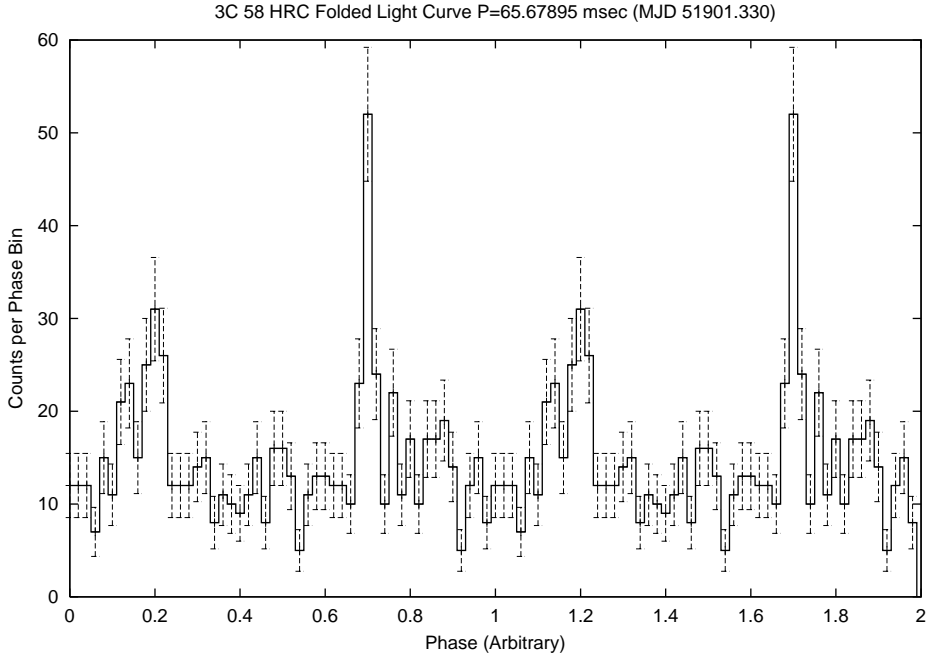} \par}

\caption{{\footnotesize The folded light curve for the HRC-S observation of
3C~58 using the 65.67895 msec period.}\label{fig:hrclc}}
\end{figure}

This list was analyzed using the FFT programs included in the IRAF
PROS package resulting in a strong signal at 30.451 Hz (corresponding
to a 32.839 msec period). A standard epoch fold around this period
gives a light curve with a single sharp pulse and a duty cycle of
about 10-15\%. We noted that the sharply peaked nature of this pulse
profile could indicate that the true period is twice as long, and
that its second harmonic could dominate the other harmonics giving
the strong FFT signal at seen at 30.451 Hz. To test this hypothesis,
we folded the data modulo a range of periods around 65.679~ms and
a over a range of period derivatives in order to maximize the signal-to-noise
of the resulting pulse profile. The best profile, with a significance
of \( \sim 6.7\sigma  \) as determined using \( \chi ^{2} \) for
a non-varying, constant model, is shown in Figure \ref{fig:hrclc}.

Here we clearly see a double pulsed signal that is slightly asymmetric,
with the two pulses about 180 degrees apart in phase angle, explaining
why the light curve for a 32.839 msec period shows one pulse. We take
the presence of two pulses of different amplitude and shape as evidence
that that we have detected a pulsar, which we designate PSR~J0205+6449,
and that the true neutron star rotation period is \( 65.67895\pm 0.00001 \)
msec (as determined on barycentric MJD 51901.330). We also find \( \dot{P}=(3.4\pm 3.5)\times 10^{-13}s/s \).
A similar analysis of the times series list was made using techniques
developed by \cite{ran01} for faint radio pulsar detection (discussed
in more detail below). As expected we obtained the same results. The
pulsed fraction for the folded light curve in Figure~\ref{fig:hrclc}
is \( \sim 21\% \). However, there is significant {}``contamination''
of the pulsed emission due to the projected emission from the extended
sub-nebula. Using the spatial model described in Section~\ref{sec:2d},
we integrated the surface brightness from the three components over
the inner 1 arc second region used for the light curve generation.
We obtained 150 counts from the pulsar (g1), 557 counts from the sub-nebula
(g2), and 8 counts of background (c3). Thus, the true pulsed fraction
could be as high as 100\%.

\section{Observations with RXTE}

To confirm the periodicity detected with \emph{Chandra}, we analyzed
an archival observation of 3C~58 carried out with the Rossi X-ray
Timing Explorer (RXTE) on 30 September, 1997 (ObsID 20259-02-01-00).
The total exposure time for this observation was 20.64 ksec, spread
over a total duration of 36.12 ksec. We analyzed data recorded with
the Proportional Counter Array (PCA) (\cite{jah96}). The PCA consists
of five identical proportional counter units (PCUs) and is sensitive
to X-rays in the energy range 2-60~keV with a time-resolution of
\( \sim 1\: \mu s \). While the PCA has no imaging capability and
collects photons across a \( 1^{\circ } \) field-of-view, its large
effective area (\( \sim 6000\: cm^{2} \)) makes it well-suited to
searching for sources of faint X-ray pulsations.

The raw telemetry packet data from the observation were analyzed using
MIT custom software optimized for RXTE analysis. We only considered
data recorded in ``GoodXenon'' mode, and included only photons which
were recorded in the top xenon layer of the PCUs and which fell into
channels 2-27 (corresponding to an approximate energy range 2-10~keV).
The data were re-sampled at \( 2^{-9}s=1.95\: ms \) time-resolution,
and periods when the source was earth-occulted or off-axis were excluded.
A correction was then applied to the data so that the binned arrival
times corresponded to barycentric dynamical time, using the \emph{Chandra}
source position.

With these data, we created a time series of 15 million points, and
applied a Fast Fourier Transform. The resulting amplitude spectrum
was searched using matched filtering techniques capable of detecting
signals with constant frequency derivatives. The search is a Fourier-domain
version of the ``acceleration'' searches used to find binary radio
pulsars and was sensitive to signals with \( \dot{f}<1.16\times 10^{-8} \)
Hz/s. Incoherent summations of 1, 2, 4 and 8 harmonics were computed
to improve sensitivity to narrow duty-cycle pulsations. Fourier amplitudes
between the raw Fourier bins were calculated in order to minimize
the ``scalloping'' of sensitivity (see \cite{ran01} for a thorough
discussion of these techniques).

The most significant candidate from the acceleration search had a
period of 65.65923(5)\textasciitilde{}msec with summed power of \( \sim 59 \)
times the local mean power level using eight harmonics. This result
confirmed the existence of periodicity found with the \emph{Chandra}
HRC\emph{.} This summed power corresponds to an overall significance
of \( \sim 6.4\sigma  \) when the number of trials searched is taken
into account. The strongest individual harmonic (the third) contained
only \( \sim 13 \) times the local mean power level, helping to explain
why earlier searches of this observation did not detect pulsations.

\begin{figure}[h]
{\centering \includegraphics{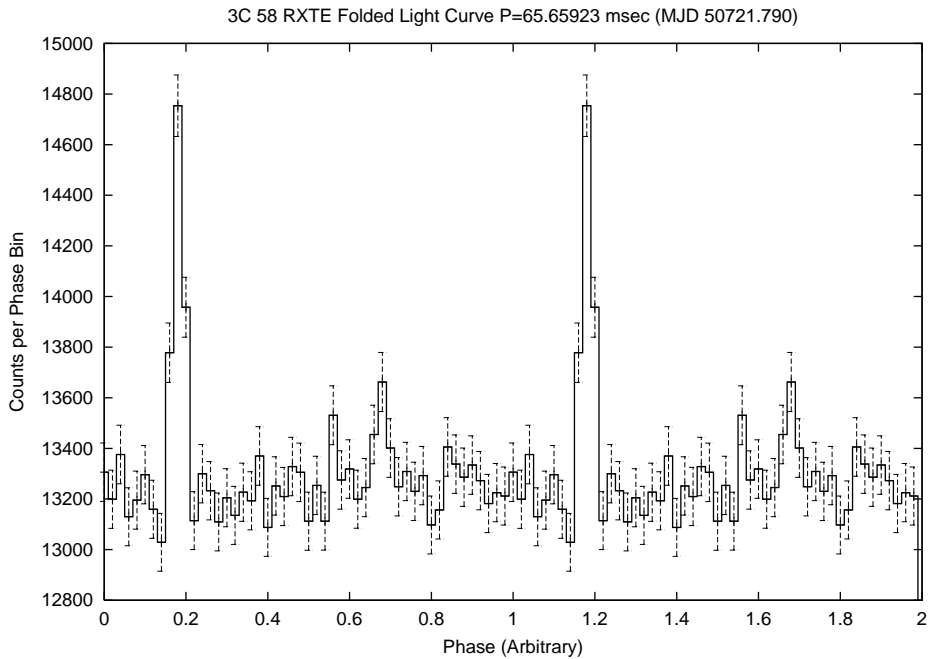} \par}

\caption{{\footnotesize Folded light curve using the RXTE observation of 3C~58
using the 65.65923 msec period.} {\small \label{fig:rxtelc}}}
\end{figure}

Once the period was known, we maximized the signal-to-noise ration
by folding various energy cuts of the raw data over a range of periods
and period derivatives centered at \( P=65.65923\: msec \) and \( \dot{P}=0.0 \).
The best pulse profile is shown in Figure \ref{fig:rxtelc} and has
a statistical significance of \( \sim 11\sigma  \) as determined
by calculating \( \chi ^{2} \) with respect to a constant model.
The best period corresponds to 65.65923(2)\textasciitilde{}msec (at
a barycentric MJD of 50721.790). The pulse shape is similar to that
seen with the HRC. However, the RXTE energy band differs from the
\emph{Chandra} band so that energy dependent features may account
for differences in the details of the light curve. Also, the RXTE
data includes many events that are not associated with the pulsar,
again making a detailed comparison with the HRC data difficult.

\section{Interpretation}

Some properties of the rotating neutron star in 3C~58 can be estimated
by applying the standard magnetic dipole model (\cite{pac67,ost69})
to the observed data. In addition to the pulse period (\( P=65.67895\: sec \)
on MJD 51901.330), we directly measured the change in period by comparing
the RXTE and \emph{Chandra} periods and obtained \( \dot{P}_{obs}=1.935(2)\times 10^{-13}s/s \).
From the spin down rate the spin down energy release is given by

\[
\dot{E}_{obs}=4\pi ^{2}I\dot{P}/P^{3}=2.6\times 10^{37}erg/s\]

where we use the canonical value for the moment of inertia (\( I=1\times 10^{45}gm\: cm^{-2}s^{-2} \))
of a \( 1.4\: M_{\odot } \) neutron star.

The surface magnetic field at the equator of the pulsar is given by:

\[
B=\sqrt{\frac{3c^{3}IP\dot{P}}{8\pi ^{2}R^{6}}}=3.6\times 10^{12}Gauss\]

The characteristic age is given by:

\[
\tau =\frac{P}{\left( n-1\right) \dot{P}}\left[ 1-\left( \frac{P_{0}}{P}\right) ^{n-1}\right] \]

where \( P_{0} \) is the initial period, and n is the braking index.
In the case of 3C~58, if we assume that \( P_{0}\ll P, \) and \( n=3 \)
), we obtain:

\[
\tau =\frac{P}{2\dot{P}}=5.38\times 10^{3}yr\]

which is longer than the historical age of 820 yr. Taking a braking
index less than 3 increases the estimated characteristic time as calculated
above, \emph{i.e.,} if the index were as small as 1.5, then the characteristic
age becomes 4 times larger. We conclude that if the historical age
is correct, then the initial spin period for this neutron star is
not negligible and we estimate its value, assuming a braking index
of 3 to be:

\[
P_{0}=P\sqrt{1-\frac{2\dot{P}\tau }{P}}=60.57\, msec\]

indicating that in this supernova, the resulting neutron star was
initially spinning slower than the current spin rate for the Crab.
The initial spin period is not a sensitive function of the assumed
braking index. For example, if the braking index is as low as n=1.5,
then the initial spin period is 60.86 msec.

\section{Discussion}

The 3C~58/J0205+6449 system is remarkably similar to the Crab Nebula/Pulsar
in several respects. While we defer a more detailed discussion of
the pulsar and compact nebula, and their relationship to the overall
energetics of 3C 58, to a future publication, we discuss here some
of the basic observational characteristics. The non-thermal spectrum
of 3C~58 indicates that most X-ray emission is from synchrotron radiation
and, since the nebula is centered on, and brightest in the vicinity
of the pulsar, we conclude that the pulsar must be the source of power.
Indeed, the \( \dot{E} \) we calculate, \( 2.6\times 10^{37}\: erg\: s^{-1} \)
can easily supply the nebular X-ray energy, \( L_{neb}=2.9\times 10^{34}erg\: s^{-1} \).
The pulsar wave-form, two narrow peaks separated 180 degrees in phase,
also indicates a non-thermal origin. The two systems, however, differ
significantly in luminosity. Nebular X-ray emission from 3C~58 is
a factor of \( \sim 1000 \) less than that of the Crab Nebula and
the pulsed X-ray luminosity of J0205+6449 is a factor of \( \sim 6000 \)
less that that of the Crab Pulsar. Although the weaker pulsed emission
might be explained by beaming, the weaker nebular emission requires
that the energy loss of the 3C~58 pulsar be considerably less than
that of the Crab. There are empirical relationships discussed in the
literature between the spin down energy of a pulsar and either the
synchrotron nebula X-ray luminosity (\emph{e.g., \cite{sew88}}) or
the total pulsar X-ray luminosity (\emph{e.g.,}\cite{bec97}). For
3C~58, both the predicted X-ray nebula luminosity and the total pulsar
luminosity are a factor of 10 or more greater than that observed. 

There have been several attempts to derive pulsar characteristics
that would fit the nebular measurements. Most inferred a pulsar having
high P and large B: \cite{sew88}, P = 550 msec B = \( 8\times 10^{13} \)
Gauss; \cite{fra93}: P = 730 msec, B=\( 1\times 10^{14} \) Gauss;
\cite{hel95}, P = 200 msec, B=\( 3\times 10^{13} \) Gauss. Others
postulated an energy source that has turned off or somehow changed
configuration (\cite{gre92,wol97}). Our observation shows, however,
that the 3C~58 pulsar is not unusual but is remarkably Crab-like.
The known age of 3C~58 leads to the conclusion that the pulsar, in
contrast to the Crab Pulsar, has not slowed much since birth, and
that the luminosity has always been low compared to the present luminosity
of the Crab Pulsar. Like the Crab Pulsar, J0205+6449 should exhibit
pulsed optical and radio emission. Now that the period is known, a
sensitive search would be worthwhile.

Although 3C~58 is commonly accepted as the remnant of SN1181, agreement
on this association is not unanimous (\cite{hua86}). \cite{bie01}
have recently observed the radio expansion of 3C~58. They suggest
that it is too slow for a free expansion of a young remnant and suggest
that the age of the remnant could be close to 5000 years, in agreement
with the characteristic age we derive for J0205+6449 assuming that
\( P_{0}\ll P \) and n=3. However, since deceleration of the remnant
might be expected, and since there are no other good candidates for
the remnant of SN1181, we contend that the first interpretation -
a low luminosity pulsar with little slowing - is more likely to be
correct.

We are left with 3C~58 containing a relatively normal pulsar (albeit
one that was born spinning

slowly) yet still being unusual with its low X-ray luminosity, radio
flux density increasing with time, and a sharp low frequency spectral
break which are all hard to understand in the context of our results.

\acknowledgements{Acknowledgments}

This work was supported in part by NASA through the Chandra \emph{}HRC
Contract, NAS 5-38248. Much of the timing analysis in this paper was
carried out on a Linux cluster at CfA funded by NSF grant PHY 9507695.
B.M.G. acknowledges the support of NASA through Hubble Fellowship
grant HST-HF-01107.01-A awarded by the Space Telescope Science Institute,
which is operated by the Association of Universities for Research
in Astronomy, Inc., for NASA under contract NAS 5-26555. P.O.S and
F.D.S acknowledge the support of NASA Contract NAS8-39073. We acknowledge
the use of the NASA Astrophysics Data System in making it easier to
review the literature and prepare our reference list.

\end{document}